\documentstyle[aps,epsf]{revtex}  

\begin{document}        

\baselineskip 14pt
\title{West Indies or Antarctica --- Direct CP Violation in B Decays
\footnote{Based on talk given at DPF99, UCLA, Jan. 1999,
reporting on work done in collaboration with
N.G. Deshpande, X.G. He, S. Pakvasa and K.C. Yang.}
}
\author{George W.S. Hou}
\address{Department of Physics,
National Taiwan University, Taipei, Taiwan 10764, R.O.C.
} 
\maketitle              

\begin{abstract}        

Discovery Voyage into The Age of: \ \ \ 
\parbox[h]{1in}{
\begin{figure}[ht]	
\centerline{\epsfxsize 0.6 truein \epsfbox{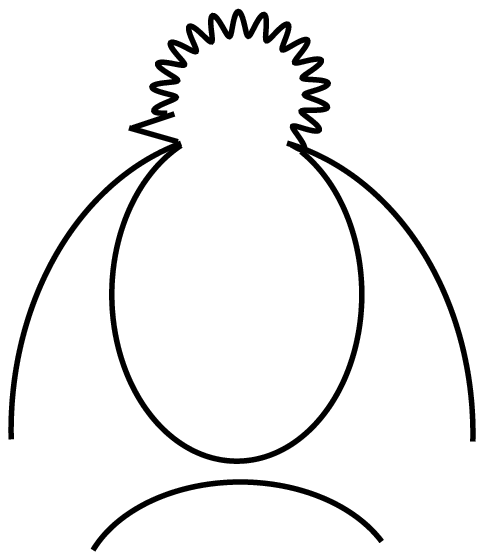}}   
\vskip -1 cm
\end{figure}
}

\
\end{abstract}   	

\section{Introduction}               

Our title clearly alludes to the story of Columbus landing in 
what he called the ``West Indies", 
which later on turned out to be part of the ``New World". 
I have substituted Antarctica in place of the ``New World",
following a quip from Frank Paige after he realized that 
I was talking all the time about {\it penguins}.
At the end of the Millennium, we are indeed on another Discovery Voyage.
We are at the dawn of observing CP violation in the B system.
The stage is the emerging penguins. 
Well, had Columbus seen penguins in {\it his} ``West Indies",
he probably would have known he was onto something really new.

The EM penguin (EMP) $B\to K^*\gamma$ (and later, $b\to s\gamma$)
was first observed by CLEO in 1993. 
Alas, it looked and walked pretty much 
according to the Standard Model (SM), and the agreement 
between theory and experiment on rates are quite good.
Perhaps the study of CP asymmetries ($a_{\rm CP}$) could reveal whether
SM holds fully.

The strong penguins (P) burst on the scene in 1997, and by now
the CLEO Collaboration has observed of order 10 exclusive modes
\cite{GaoWuerth},
as well as the surprisingly large inclusive $B \to \eta^\prime + X_s$ mode.
The $\eta^\prime K^+$, $\eta^\prime K^0$ and $K^+\pi^-$ modes are rather
robust, but the $K^0\pi^+$ and $K^+\pi^0$ rates shifted when CLEO II data
were recalibrated in 1998 and part of CLEO II.V data were included. 
The $\omega K^+$ and $\omega \pi^+$ modes are still being reanalyzed. 
The nonobservation, so far, of the $\pi^+\pi^-$, $\pi^+\pi^0$ 
and $\phi K^+$ modes are also rather stringent. 
The observation of the $\rho^0\pi^+$ mode was 
announced in January this year,
while the observation of the $\rho^\pm\pi^\mp$ and $K^{*+}\pi^-$ modes 
were announced in March. 
CLEO II.V data taking ended in February.
With 10 million or so each of charged and neutral B's, 
new results are expected by summer and certainly by winter.
Perhaps the first observation of direct CP violation could be reported soon.

With BELLE and BABAR turning on in May, together with the CLEO III
detector upgrade --- all with $K/\pi$ separation (PID) capability! ---
we have a three way race for detecting and eventually disentangling 
{\it direct} CP violation in charmless B decays. 
We expect that, during 1999--2002,  
the number of observed modes may increase to a few dozen, 
while the events per mode may increase from 10--70 
to $10^2$--$10^3$ events for some modes, 
and sensitivity for direct CP asymmetries would go from
the present level of order 30\% down to 10\% or so.
It should be realized that
{\it the modes that are already observed} ($b\to s$)
{\it should be the most sensitive probes.}

Our first theme is therefore: 
{\it Is Large $a_{\rm CP}$ possible in $b\to s$ processes?}
and, {\it If so, Whither New Physics?}
However, as an antidote against the rush into the brave New World,
we point out that the three observed $K\pi$ modes
may indicate that the ``West Indies" interpretation is still correct so far.
Our second subject would hence be
{\it Whither EWP? Now!?}
That is, we will argue for the intriguing possibility that 
perhaps we already have some indication for the electroweak penguin (EWP).

It is clear that 1999 would be an exciting landmark year in B physics.
So, work hard and come party at the end of the year/century/millennium
celebration called ``Third International Conference on B Physics and CP
Violation", held December 3-7 in Taipei \cite{BCP3}.

\section{Is Large CP Violation Possible/Whither New Physics?}

We shall motivate the physics and give some results 
that have not been presented before,
but refer to more detailed discussions that
can be found elsewhere \cite{largeCP,HHY}.

Our interests were stirred by a {\it rumor} in 1997 that CLEO had
a very large $a_{\rm CP}$ in the $K^+\pi^-$ mode. 
The question was: {\it How to get large $a_{\rm CP}$?}
With short distance (Bander-Silverman-Soni \cite{BSS}) 
rescattering phase from penguin,
the CP asymmetry could reach its maximum of order 10\% around the
presently preferred $\gamma \simeq 64^\circ$.
Final state $K\pi \to K\pi$ rescattering phases 
could bring this up to 30\% or so,
and would hence mask New Physics.
But a 50\% asymmetry seems difficult.
New Physics asymmetries in the $b\to s\gamma$ process \cite{WW} and
$B\to \eta^\prime  + X_s$ process \cite{HT} are typically of order 10\%,
whereas asymmetries for penguin dominant $b\to s$ transitions
are expected to be no more than 1\%.

The answer to the above challenge is to {\it hit SM at its weakest!}
\begin{itemize}
\item {\it Weak Spot of Penguin}: Dipole Transition

\vskip-0.3cm 
\hskip0.8cm
\parbox[h]{1.2in}{
\begin{figure}[ht]	
\centerline{\epsfxsize 1.2 truein \epsfbox{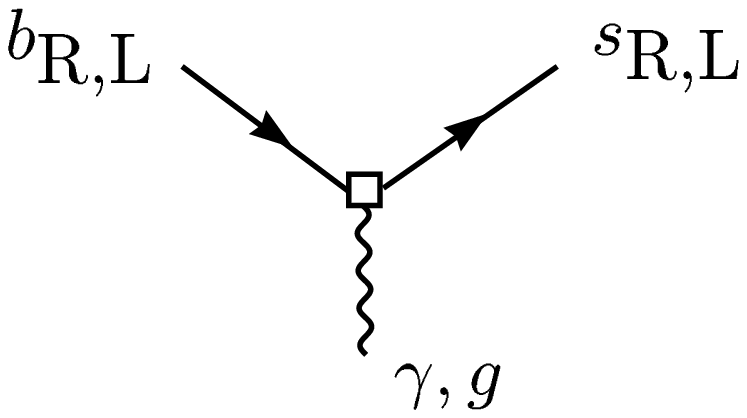}}   
\vskip -0.25 cm
\end{figure}
}\hskip1.3cm 
\parbox[h]{4in}{
 $F_1 \,  (q^2\gamma_\mu - q_\mu \not{\! q}) L 
+ \underline{\underline{F_2}} \, i\sigma_{\mu\nu} q_\nu m_b R$
\vskip -0.4 cm
}

Note that these two terms are at same order in 
$q/M_W$ and $m_b/MW$ expansion.
The effective ``charge" is $F_1 q^2$ which vanishes when
the $\gamma$ or $g$ goes on-shell, hence, 
only the $F_2$ dipole enters $b\to s\gamma$ and $b\to sg$ transitions.
It is an SM quirk due to the GIM mechanism
that $\vert F_1\vert \gg \vert F_2 \vert$
(the former becoming $c_{3-6}$ coefficients in 
usual operator formalism for gluonic penguin).
Hence one usually does not pay attention to the subdominant
$F_2^g$ which goes into the variously called $c_8$, $c_g$, or
$c_{11}$ coefficients. 
In particular, $b\to sg$ rate in SM is only of order 0.2\%.
But if New Physics is present, 
having $\delta F_2 \sim \delta F_1$ is natural,
hence the gluonic dipole could get greatly enhanced.
While subject to $b\to s\gamma$ constraint,
this could have great impact on $b\to sg^* \to sq\bar q$ process.

\item {\it Blind Spot of Detector!}

Because $b\to sg$ leads to 
{\it jetty, high multiplicity} $b\to s$ transitions

\vskip-0.3cm 
\hskip0.8cm
\parbox[h]{2.2in}{
\begin{figure}[ht]	
\centerline{\epsfxsize 2.2 truein \epsfbox{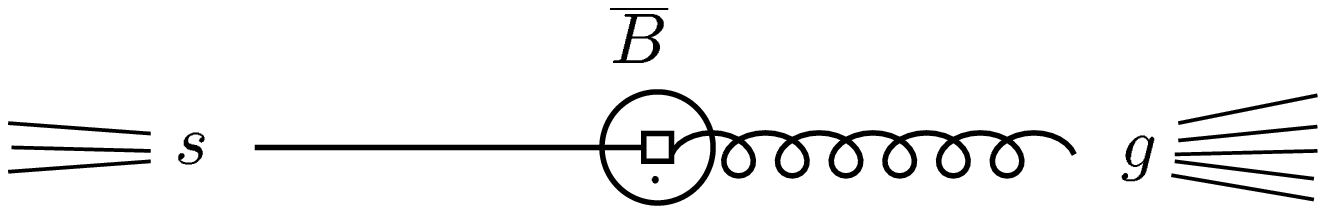}}   
\vskip -0.2 cm
\end{figure}
}
\hskip0.9cm 
\parbox[h]{3.5in}{
Hide easily in dominant $b\to c\to s$ sequence!
\vskip -0.5 cm
}

At present, 5--10\% could still easily be allowed.
The semileptonic branching ratio and charm counting deficits,
and the strength of $B\to \eta^\prime + X_s$ rate provide
circumstantial {\it hints} that $b\to sg$ could be more than a few percent.

\item  {\it Unconstrained new CP phase} via $b_R \to s_L$

If enhanced by New Physics, $F_2^g$ is likely to carry a New Phase

\vskip-0.27cm 
\hskip0.8cm
\parbox[h]{1.2in}{
\begin{figure}[ht]	
\centerline{\epsfxsize 1.2 truein \epsfbox{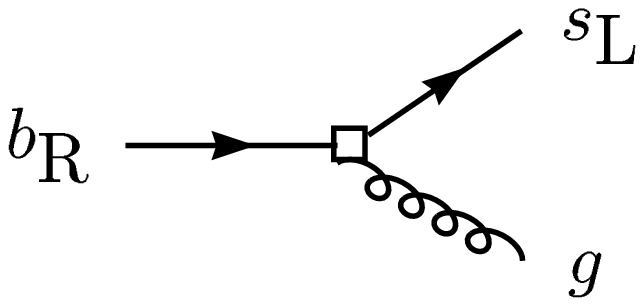}}   
\vskip -0.27 cm
\end{figure}
}
\hskip0.9cm 
\parbox[h]{4in}{
Phase of $b_R$ not probed by $V_{\rm CKM}!$
\vskip -0.4 cm
}

However, one faces a severe constraint from $b\to s\gamma$. 
For example it rules out the possibility of $H^+$ as source of enhancement.
But as Alex Kagan \cite{Kagan} taught me at last DPF meeting in Minnesota,
the constraint can be evaded
if one has sources for radiating $g$ but not $\gamma$.

\item Uncharted territory of Nonuniversal Squark Masses

SUSY provides a natural possibility via gluino loops:

\vskip-0.35cm 
\hskip0.9cm
\parbox[h]{1in}{
\begin{figure}[ht]	
\centerline{\epsfxsize 1.45 truein \epsfbox{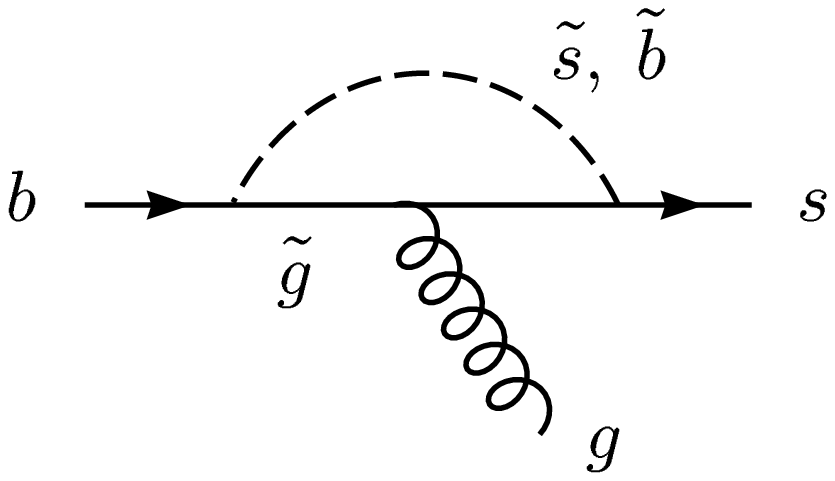}}   
\vskip -0.8 cm
\end{figure}
}
\hskip1.3cm 
\parbox[h]{4in}{
Need flavor violation in $\tilde d_j$
\vskip 0.3 cm
}

The simplest being a $\tilde s$--$\tilde b$ mixing model \cite{HH,CHH}.
Since the first generation down squark is not involved,
one evades all low energy constraints.
This is a New Physics CP model tailor-made for $b\to s$ transitions.
\end{itemize}

With the aim of generating huge CP asymmetries,
we can now take $b\to sg \sim 10\%$
and study $b\to sq\bar q$ transitions 
at both inclusive and exclusive level \cite{HHY}.
In both we have used operator language.
One needs to consider the tree diagram, which carries
the CP phase $\gamma \equiv {\rm arg}\,(V_{ub}^*)$;
the standard penguin diagrams, which contain 
short distance rescattering phases;
the enhanced $bsg$ dipole (SUSY loop induced) diagram;
finally, diagrams containing $q\bar q$ loop insertions to the 
gluon self-energy which are needed to maintain unitarity and
consistency to order $\alpha_S^2$ in rate differences \cite{GH}.

At the inclusive level, 
one finds a ``$b\to sg$ pole" at low $q^2$ which 
reflects the jetty $b\to sg$ process
that is experimentally hard to identify.
Destructive interference is in general
needed to allow the $b\to sq\bar q$ rate to be comparable to SM.
But this precisely facilitates the generation of large $a_{\rm CP}$s!
More details such as figures can be found in \cite{largeCP,HHY}.
Dominant rate asymmetry comes from large $q^2$ of the virtual gluon.
To illustrate this, Table I gives inclusive BR (arbitrarily cutoff at
$q^2 = 1$ GeV$^2$) and $a_{\rm CP}$ for SM and 
for various new CP phase $\sigma$ valus, 
assuming $b\to sg$ rate of order 10\%. 
One obtains SM-like branching ratios for
$\sigma \simeq 145^\circ$, and $a_{\rm CP}$ also seem to peak.
This becomes clearer in Table II where we give
the results for $q^2 > 4m_c^2$,
where $c\bar c \to q\bar q$ (perturbative) rescattering is fully open.
We see that 20--30\% asymmetries are achieveable.
This provides support for findings in exclusive processes.

Exclusive two body modes are much more problematic.
Starting from the operator formalism as in inclusive,
we set $N_C = 3$, take $q^2 \sim m_b^2/2$
and try to {\it fit observed {\rm BRs}}
with $b\to sg \simeq 10\%$.
We then find the $a_{\rm CP}$ preferred by present rate data.
One finds that, analogous to the inclusive case, 
destructive interference is needed
and in fact provides a mechanism to 
suppress the pure penguin $B\to \phi K^+$ mode to satisfy CLEO bound.
For the $K^+\pi^-$ and $K^0\pi^+$ modes which are P-dominated,
one utilizes the fact that the matrix element 
\[
\langle O_6\rangle \propto {m_K^2 (m_B^2 - m_\pi^2)
                            \over (\underline{\underline{m_s}} + m_u)
                                  (m_b - m_u)}
\]
could be enhanced by low $m_s$ values (of order 100--120 MeV)
to raise $K\pi/\phi K$,
which at same time leads to near degeneracy
of $K^+\pi^-$ and $K^0\pi^+$ rates.
The upshot is that one finds rather large CP asymmetries,
i.e. $a_{\rm CP} \sim $ 35\%, 45\% and 55\% for
$K^0\pi^+$, $K^+\pi^-$ and $\phi K^+$ modes, respectively,
and all of the same sign.
Such pattern cannot be generated by SM,
with or without rescattering.
We expect such pattern to hold true for many $b\to s$ modes.

\begin{table}[bht]
\caption{Inclusive BR (in $10^{-3}$)/$a_{\rm CP}$ (in \%) 
for SM and for $c_8 = 2 e^{i\sigma}$.} 
\begin{tabular}{cccccccc}
&SM &$\sigma$ =\ \ \ \ \ \  0 \ \ \ \ \ & $\frac{i\pi}{4}$ &
 $\frac{i\pi}{2}$ & $\frac{i3\pi}{4}$ & $i\pi$  \\
\hline $b\to s \bar d d$ & 2.6/0.8   & \ \ \ \ \ \ 8.5/0.4 & 7.6/3.4 &
5.2/6.5 & 2.9/8.1 & 1.9/0.5 \\ 
\hline $b\to s \bar u u$ & 2.4/1.4   & \ \ \ \ \ \ \ 8.1/-0.2 & 7.5/2.6 &
5.5/5.6 & 3.2/8.1 & 2.0/3.5 \\ 
\hline $b\to s \bar s s$ & 2.0/0.9 & \ \ \ \ \ \ 6.9/0.4 & 6.2/3.2 &
4.4/6.0 & 2.6/7.1 & 1.8/0.4 
\end{tabular}
\vskip -0.2cm
\end{table}
\begin{table}[bht]
\caption{Inclusive BR (in $10^{-3}$)/$a_{\rm CP}$ (in \%) for SM
and for $c_8 = 2 e^{i\sigma}$ above the $4m_c^2$ threshold.}
\begin{tabular}{cccccccc}
&SM &$\sigma$ =\ \ \ \ \ \  0 \ \ \ \ \ & $\frac{i\pi}{4}$ &
 $\frac{i\pi}{2}$ & $\frac{i3\pi}{4}$ & $i\pi$  \\
\hline
 $b\to s \bar d d$
 & 1.4/0.5 & \ \ \ \ \ \ 3.1/0.3&2.8/8.2 &1.9/16.8 & 1.0/22.9 & 0.6/0.7
 \\ \hline 
 $b\to s \bar u u$
 &1.3/4.6  & \ \ \ \ \ \ 3.0/1.1&2.7/9.0&1.9/17.9&1.1/26.2&0.6/2.8
 \\ \hline 
 $b\to s \bar s s$
 &0.5/0.5 & \ \ \ \ \ \ 1.1/0.3 &1.0/7.1&0.7/14.8&0.3/21.6&0.2/0.9
\end{tabular}
\end{table}

We have left out the prominent $B\to \eta^\prime K$ modes
from our discussion largely because the anomaly contribution

\vskip-0.35cm 
\hskip0.9cm
\parbox[h]{1.45in}{
\begin{figure}[ht]	
\centerline{\epsfxsize 1.45 truein \epsfbox{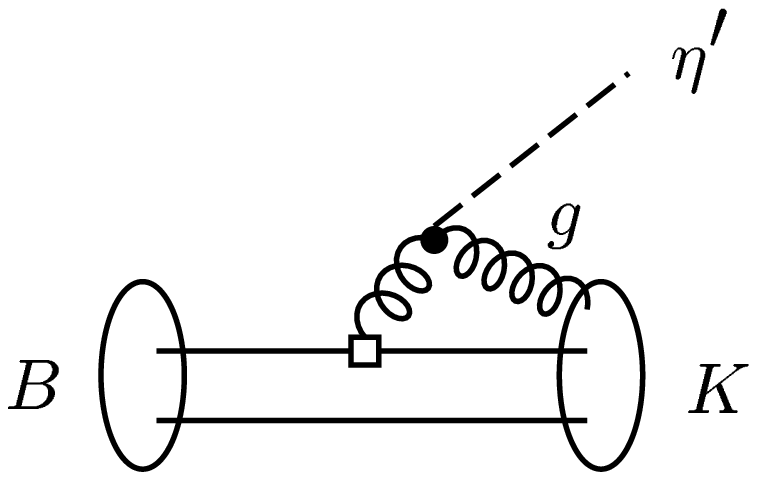}}   
\vskip -0.3 cm
\end{figure}
}
\hskip1.3cm 
\parbox[h]{3in}{
Not quite included at present!
\vskip -0.9 cm
}

\noindent To compute such diagrams, one needs to know the $\vert \bar
sgq\rangle$ Fock component of the $K$ meson!
This may be at the root of the 
rather large size of $B\to \eta^\prime K$ mode.

\section{Whither EWP? Now!?}

Before we get carried away by the possibility of 
large CP asymmetries from New Physics, 
there is one flaw (or two?)
that emerged after summer 1998.
Because of P-dominance which is
certainly true in case of enhanced $b\to sg$,
$K^+\pi^0$ is only half of $K^+\pi^- \simeq K^0\pi^+$. 
The factor of 1/2 comes from
$A^P_{K^+\pi^0} \sim \frac{1}{\sqrt 2} A^P_{K^+\pi^-}$,
which is just an isospin Clebsch factor that
originates from the $\pi^0$ wave function.
Although this seemed quite reasonable from 1997 data
where $K^+\pi^0$ mode was not reported,
a crisis emerged in summer 1998 when CLEO updated
their results for the three $K\pi$ modes.
They found \cite{Kpi0} $K^+\pi^0 \simeq K^+\pi^- \simeq K^0\pi^+$
instead!

Curiously,
$A^T_{K^+\pi^0} \sim \frac{1}{\sqrt 2} A^T_{K^+\pi^-}$ also,
which cannot change the situation.
In any case the expectation that $\vert T/P\vert \sim 0.2$
cannot make a factor of 2 change by interference.
Miraculously, however,
this could be the first indication of 
the last type of penguin, the EWP.

The yet to be observed EWP (electroweak penguin), 
namely $b\to sf\bar f$, occurs by
$b\to s\gamma^*,\ Z^*$ followed by $\gamma^*,\ Z^* \to f\bar f$.
The strong penguin oftentimes obscure the $b\to sq\bar q$ case
(or so it is thought),
and to cleanly identify the EWP one has to search for
``pure" EWP modes such as $B_s \to \pi\eta$, $\pi\phi$
which are clearly rather far away.
One usually expects the $B\to K^{(*)}\ell^+\ell^-$ mode
to be the first EWP to be observed, 
which is still a year or two away,
while clean and purely weak penguin $B\to K^{(*)}\nu\bar\nu$
is rather far away.

With the hint from $K^+\pi^0 \simeq K^+\pi^- \simeq K^0\pi^+$,
however, and putting back on our SM hat,
we wish to establish the possibility that
EWP may be operating behind the scene already \cite{DHHP}.
It should be emphasized that, unlike the gluon, 
the $Zf\bar f$ coupling depends on isospin,
and can in principle break the isospin factor of 1/2 mentioned earlier.

\begin{figure}[htb]	
\centerline{
\epsfxsize 3.2 truein \epsfbox{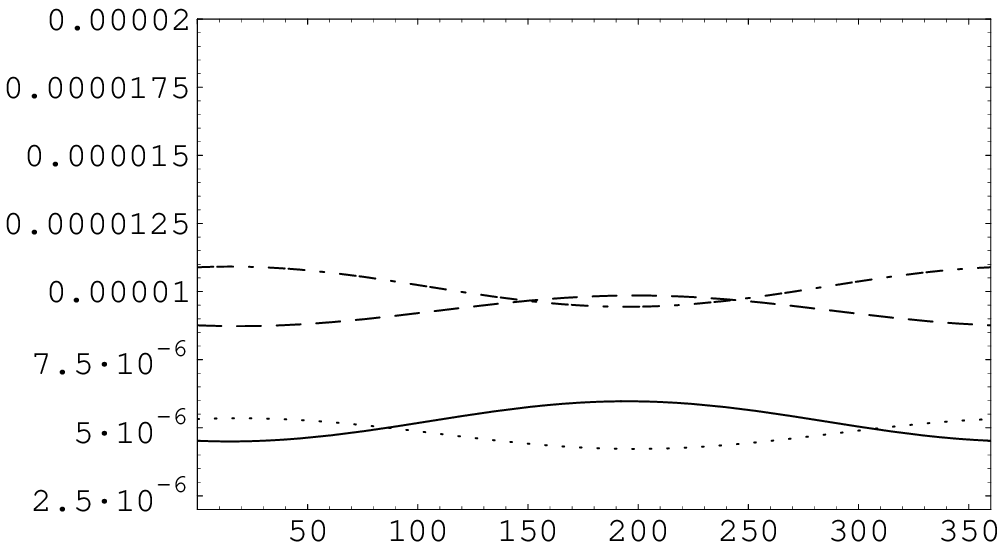} \ \ 
\epsfxsize 3.2 truein \epsfbox{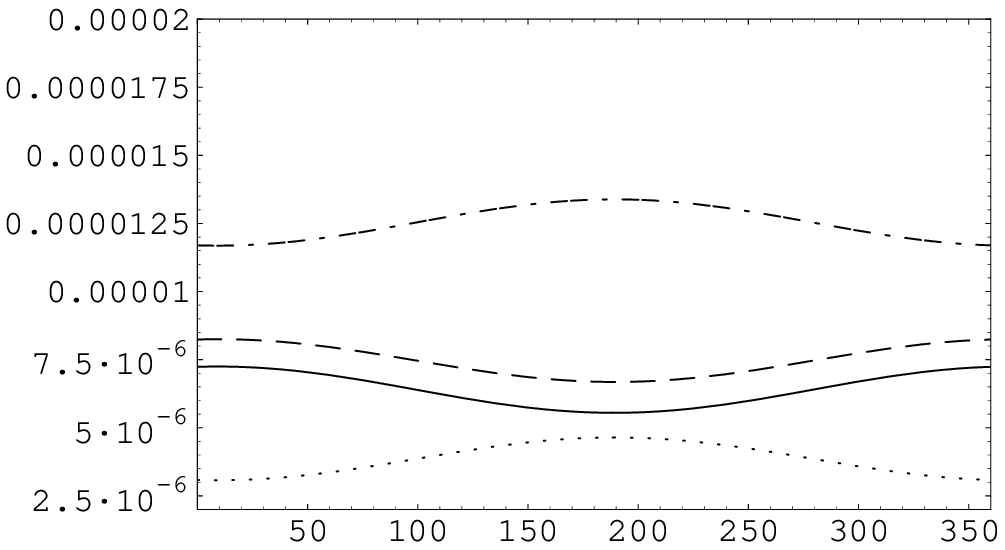} 
}
\vskip -.2 cm
\caption[]{
\label{br64}
\small BR$(B\to K\pi)$ vs. $\delta$ for $\gamma = 64^\circ$
without or with EWP  ($N= 3$, $m_s = 200$ MeV).
Solid, dot-dashed, dashed and dotted lines $ \equiv 
B^+\to K^+\pi^0,\ K^0 \pi^+$ and 
$B^0 \to K^+\pi^-,\ K^0\pi^0$.
}
\end{figure}

We first show that simple $K\pi \to K\pi$ rescattering cannot
change drastically the factor of two.
From Fig. 1(a), where we have adopted $\gamma = 64^\circ$ 
from current ``best fit" to CKM matrix \cite{Parodi},
one clearly sees the factor of 2 between
$K^+\pi^-$ and $K^+\pi^0$.
We also not that rescattering, as parametrized by the 
phase difference $\delta$ between I = 1/2 and 3/2 amplitudes,
is only between $K^+\pi^0 \leftrightarrow K^0 \pi^+$
and $K^+\pi^- \leftrightarrow K^0\pi^0$.
When we put in the EWP contribution,
at first sight it seems that the effect is drastic.
On closer inspection at $\delta = 0$,
it is clear that the EWP contribution to
$K^0 \pi^+$ and $K^+\pi^-$ modes are small,
but is quite visible
for $K^+\pi^0$ and $K^0\pi^0$ modes.
This is because the $K^+\pi^0$ and $K^0\pi^0$ modes suffer 
from $1/\sqrt 2$ suppression in amplitude
because of $\pi^0$ wave function.
However, it is precisely these
modes which pick up a sizable $Z$ penguin contribution 
via the $\pi^0$ 
(the strength of $c_9$ is roughly a quarter of $c_4$ and $c_6$).
As one dials $\delta$, $K^+\pi^0 \leftrightarrow K^0 \pi^+$
and $K^+\pi^- \leftrightarrow K^0\pi^0$ rescattering 
redistributes this EWP impact and leads to the
rather visible change in Fig. 1(b).
We notice the remarkable result that 
the EWP reduces $K^+\pi^-$ rate slightly but
raises the $K^+\pi^0$ rate considerably,
such that the two modes become rather close.
We have to admit, however, to something that we have sneaked in.
To enhance the relative importance of EWP,
we had to suppress the strong penguin effect. 
We have therefore employed a much heavier $m_s = 200$ MeV 
as compared to 100--120 MeV employed previously
in New Physics case.
Otherwise we cannot bring $K^+\pi^-$ and  $K^+\pi^0$ rates 
close to each other.

\begin{figure}[htb]	
\centerline{
\epsfxsize 3.2 truein \epsfbox{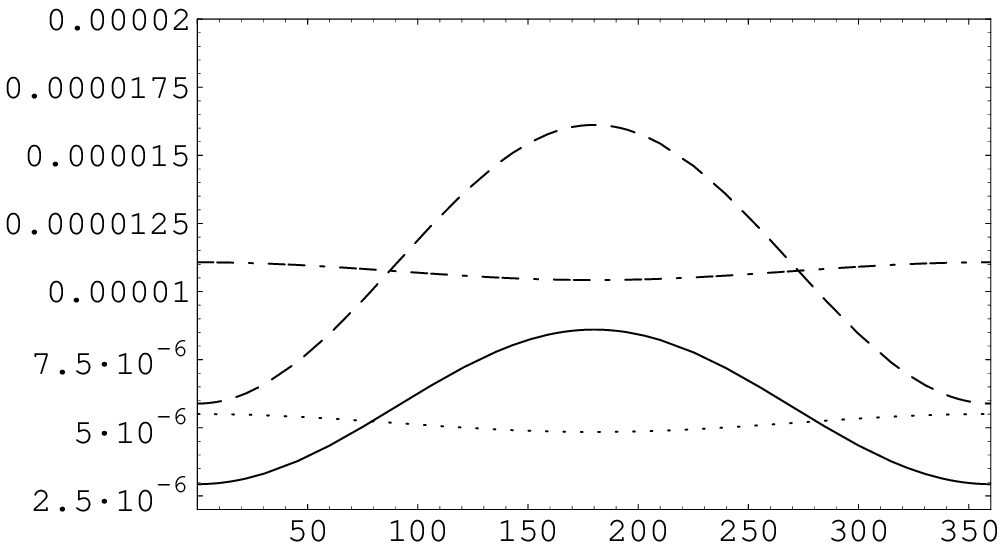} \ \ 
\epsfxsize 3.2 truein \epsfbox{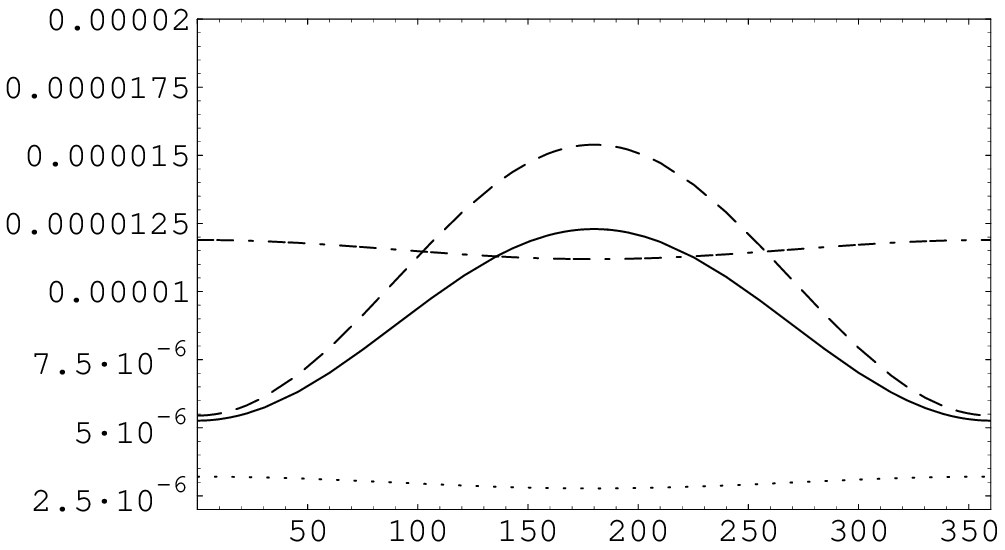} 
}
\vskip -.2 cm
\caption[]{
\label{brgamma}
\small As in Fig. 1 but vs. $\gamma$ for $\delta = 0$.
}
\end{figure}

Having brought $K^+\pi^-$ and $K^+\pi^0$ modes closer,
the problem now is that $K^0\pi^+$ lies above them,
and the situation becomes worse for large rescattering.
To remedy this,
we play with the phase angle $\gamma$ which tunes the weak phase of
the tree contribution T.
Setting now $\delta = 0$, again we start without EWP in Fig. 2(a).
The factor of two between $K^+\pi^-$ and $K^+\pi^0$
is again apparent.
Dialing $\gamma$ clearly changes T-P interference.
For $\gamma$ in first quadrant one has destructive interference,
which becomes constructive in second quadrant.
This allows the $K^+\pi^-$ mode to become larger than the
pure penguin $K^0\pi^+$ mode,
which is insensitive to $\gamma$.
However,
nowhere do we find a solution where 
$K^+\pi^0 \simeq K^+\pi^- \simeq K^0\pi^+$
is approximately true.
There is always one mode that is split away from the other two.

Putting in EWP, as shown in Fig. 2(b), the impact is again quite visible.
As anticipated,
the $K^+\pi^-$ and $K^+\pi^0$ modes come close to each other.
Since their $\gamma$ dependence is quite similar,
one finds that for $\gamma \sim 90^\circ$--$130^\circ$,
the three observed $K\pi$ modes
come together as close as one can get, 
and are basically consistent with errors allowed by data.
Note that $K^+\pi^0$ is never larger than $K^+\pi^-$.

We emphasize that a large rescattering phase $\delta$
would destroy this achieved approximate equality, as can be seen 
from Fig. 3, where we illustrate $\delta$ dependence 
for $\gamma = 120^\circ$. It seems that
$\delta$ cannot be larger than $50^\circ$ or so.

\begin{figure}[htb]	
\centerline{
\epsfxsize 3.2 truein \epsfbox{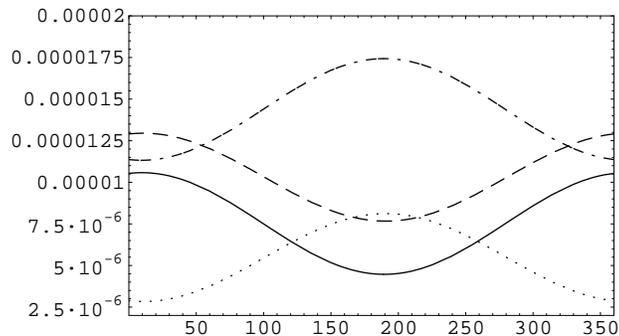} 
}
\vskip -.2 cm
\caption[]{
\label{br120}
\small As in Fig. 1 but vs. $\delta$ for $\gamma = 120^\circ$.
}
\end{figure}

\begin{figure}[htb]	
\centerline{
\epsfxsize 3.2 truein \epsfbox{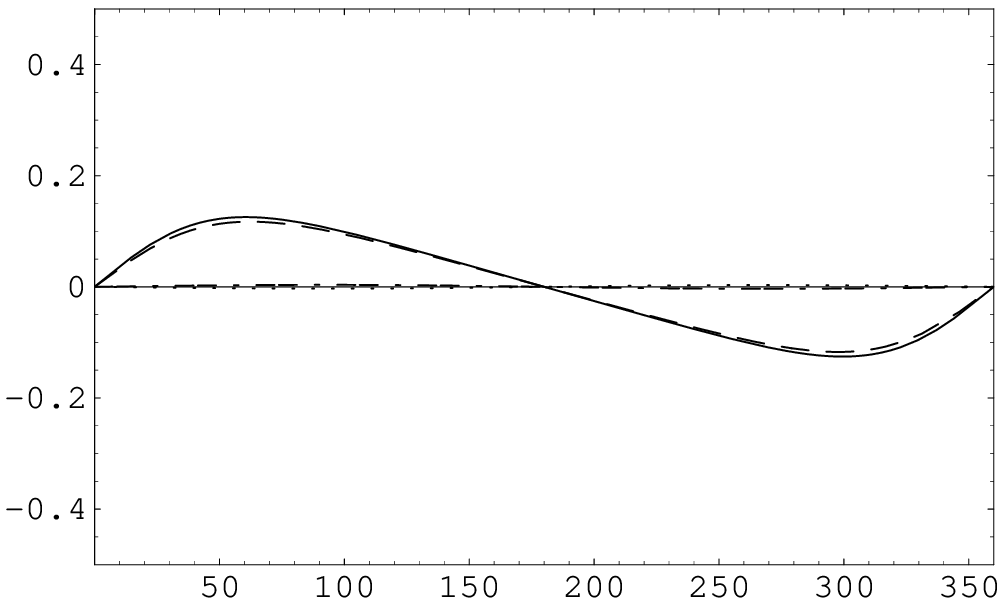} \ \ 
\epsfxsize 3.2 truein \epsfbox{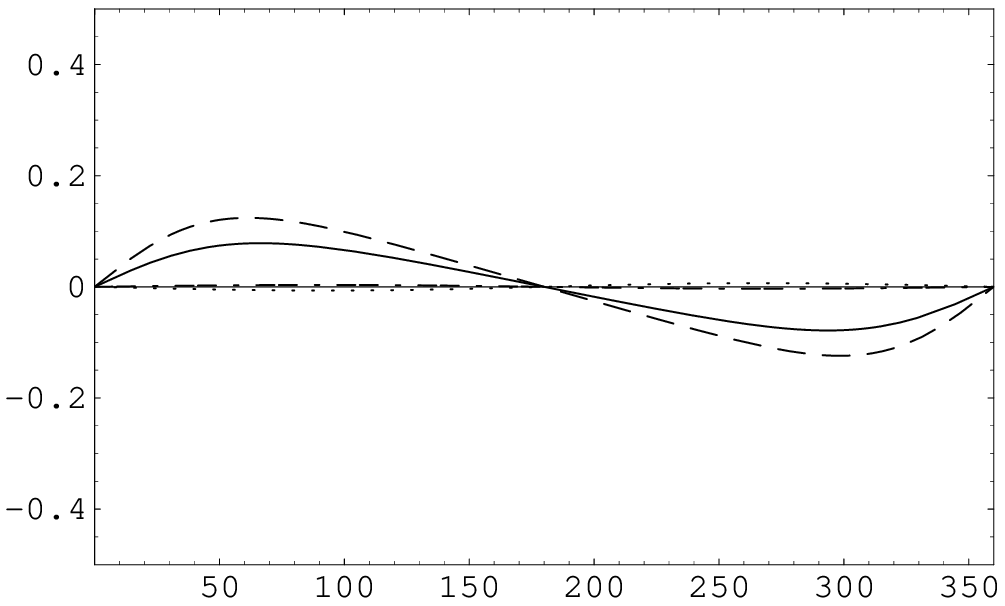} 
}
\vskip -.2 cm
\caption[]{
\label{asygamma}
\small Asymmetry vs. $\gamma$ for $\delta = 0$.
}
\end{figure}

As a further check of effect of the EWP,
we show the results for $\delta = 0$ in Fig. 4.
In absence of rescattering,
the change in rate (enhancement) for $K^+\pi^0$ mode from adding EWP
is reflected in a dilution of the asymmetry,
which could serve as a further test.
This, however, depends rather crucially on absence of rescattering.
Once rescattering is included, it would be
hard to distinguish the impact of EWP from CP asymmetries.
However, even with rescattering phase,
the $\gamma$ dependence of CP asymmetries can
easily distinguish between the two solutions of
$\gamma \sim 120^\circ$ and $240^\circ$,
as illustrated in Fig. 5, where EWP effect is included.
From our observation that a large $\delta$ phase would 
destroy the near equality of the three observed $K\pi$ modes that 
we had obtained,
we find that $a_{\rm CP} < 20\%$ even with presence of rescattering
phase $\delta$.

\begin{figure}[htb]	
\centerline{
\epsfxsize 3.2 truein \epsfbox{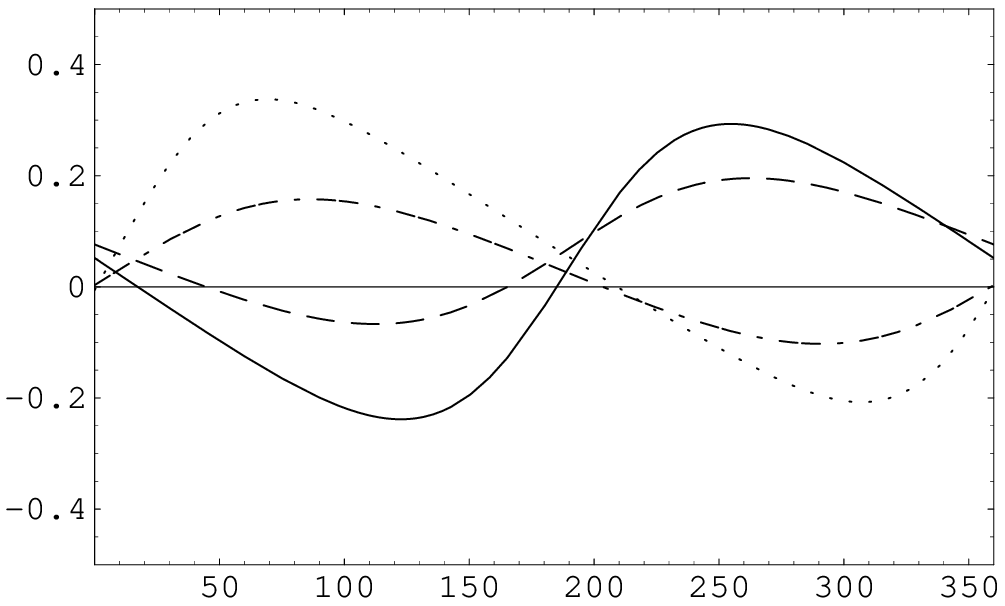} \ \ 
\epsfxsize 3.2 truein \epsfbox{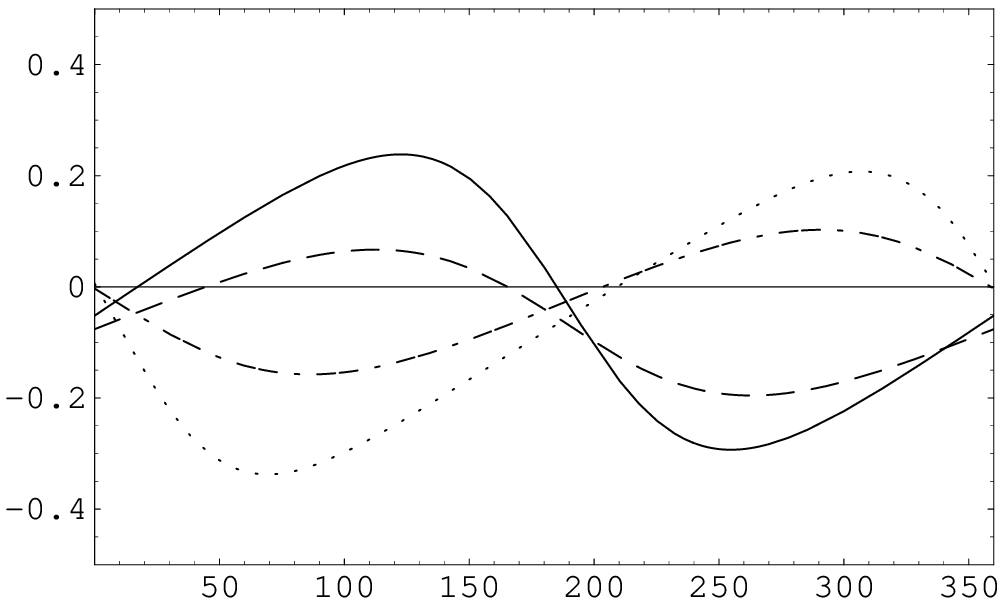} 
}
\vskip -.2 cm
\caption[]{
\label{asy}
\small Asymmetry vs. $\delta$ for $\gamma = 120^\circ$ and $240^\circ$.
}
\end{figure}

It should be emphasized that
the $\gamma$ value we find necessary to have
$K^+\pi^- \simeq K^0\pi^+$ is in a different quadrant than
the present best `fit" result of $\gamma \sim 60^\circ$--$70^\circ$.
In particular, the sign of $\cos\gamma$ is preferred to
be negative rather than positive.
An extended analysis \cite{gamma} to
$\pi\pi$, $\rho\pi$ and $K^*\pi$ modes
confirm this assertion.
Intriguingly,
the size of $\rho^\pm\pi^\mp$ and $K^{*+}\pi^-$ \cite{GaoWuerth} was 
anticipated via this $\gamma$ value.
Perhaps hadronic rare B decays
can provide information on $\gamma$,
and present results seem to be at odds with CKM fits \cite{Parodi} to 
$\varepsilon_K$, $\vert V_{ub}/V_{cb}\vert$, $B_d$
mixing,  and in particular the $B_s$ mixing bound, which
rules out $\cos\gamma < 0$.

\section{Conclusion}

\underline{\bf B}e prepared for \underline{\bf CP} Violation!!

We first illustrated
the possibility of having $a_{\rm CP} \sim 30\%$--$50\%$ 
from New Physics in
{\it already observed modes}, such as
$K\pi$, $\eta^\prime K$, and $\phi K$ mode when seen.
Our ``existence proof" was the possibility of
enhanced $b\to sg$ dipole transition,
which from SUSY model considerations one could have
a new CP phase carried by $b_R$.
Note that this is just an illustration. 
We are quite sure that Nature is smatter.

We then made an about-face 
and went back to SM,
and pointed out that the EWP may have already shone through
the special ``slit" of $K^+\pi^0 \simeq K^+\pi^- \simeq K^0\pi^+$,
where we inferred that
$\gamma \sim 90^\circ$--$130^\circ$ is preferred,
which implies that $\cos\gamma < 0$,
contrary to current CKM ``fit" preference.

We hope we have illustrated the versatility of 
rare B decays, that they can open windows on both New Physics and SM.
The next 5 years should be a very rewarding period!

\end{document}